\documentclass[aps,prl,twocolumn,reprint,showpacs,groupedaddress]{revtex4-1}
\usepackage{amssymb,amsmath,amsfonts} \usepackage{epsfig,graphicx}
\usepackage{txfonts}

\usepackage{latexsym}
\usepackage{stmaryrd}
\usepackage[table,dvipsnames]{xcolor}
\usepackage[colorlinks=true,citecolor=PineGreen,linkcolor=PineGreen,urlcolor=black]{hyperref}%
\usepackage{bm}
\usepackage{bbm}
\usepackage{color}
\usepackage[english]{babel}
\usepackage{subfigure}
\usepackage{natbib}
\usepackage{appendix}

\newcommand{\adag}{a^{\dagger}}
\newcommand{\adaga}{a^{\dagger}a}
\newcommand\ket[1]{\left|\textstyle{#1}\right\rangle}
\newcommand\bra[1]{\left\langle\textstyle{#1}\right|}

\makeatletter

\begin{document}
\title{Finite-component dynamical quantum phase transitions}
\author{Ricardo Puebla}
\email{r.puebla@qub.ac.uk}
\affiliation{Centre for Theoretical Atomic, Molecular and Optical Physics, School of Mathematics and Physics, Queen's University Belfast, Belfast BT7 1NN, United Kingdom}
\date{\today}
\begin{abstract}
  Phase transitions have recently been formulated in the time domain of quantum many-body systems, a phenomenon  dubbed  dynamical quantum phase transitions (DPTs), whose phenomenology is often divided in two types. One refers to distinct phases according to long-time averaged order parameters, while the other is focused on the non-analytical behavior emerging in the rate function of the Loschmidt echo. Here we show that such DPTs can be found in systems with few degrees of freedom, i.e. they can take place without resorting to the traditional thermodynamic limit. We illustrate this by showing the existence of the two types of DPTs in a quantum Rabi model ---a system involving a spin-$\frac{1}{2}$ and a bosonic mode. The dynamical criticality appears in the limit of an infinitely large ratio of the spin frequency with respect to the bosonic one. We determine its dynamical phase diagram and study the long-time averaged order parameters, whose semiclassical approximation yields a jump at the transition point. We find the critical times at which the rate function becomes non-analytical, showing its associated critical exponent as well as the corrections introduced by a finite frequency ratio.  Our results open the door for the study of DPTs without the need to scale up the number of components, thus allowing for their investigation in well controllable systems.
 \end{abstract}

\maketitle

{\em Introduction.--} Quantum phase transitions (QPTs) are key for the understanding of the different collective behavior exhibited by complex systems at the quantum level~\cite{Sachdev}. Such QPTs occur at zero temperature, i.e., they take place in the quantum mechanical ground state of the Hamiltonian describing the system. As in the classical or thermal phase transitions~\cite{Huang}, the change of phase is typically accompanied by a singular behavior of relevant quantities, as for example, a diverging correlation length at the critical point of continuous phase transition~\cite{Sachdev,Huang}. However, such a singular or {\em critical} behavior, either of classical or quantum nature, is only expected in the thermodynamic limit of many components.  This naturally poses a challenge for the observation of quantum critical phenomena. In spite of the tremendous experimental progress made during the last decades, to control, manipulate, and isolate a truly quantum many-body system is still a herculean task, although exceptions are notable~\cite{Friedenauer:08,Kim:10,Islam:13,Jurcevic:14,Lanyon:17,Neyenhuis:17,Jurcevic:17,Zhang:17b,Keesling:19}.

However, and quite remarkably, scaling up the number of components is not the unique route towards critical phenomena. Indeed, QPTs have been found in systems made of a few number of interacting subsystems, i.e., finite-component systems~\cite{Hwang:15,Hwang:16,Puebla:16,Larson:17,Liu:17,Shen:17,Wang:18,Hwang:18,Peng:19,Zhu:20,Felicetti:20}. In these systems, rather than in the traditional thermodynamic limit, {\em criticality} appears in a suitable parameter limit while keeping fixed the number of components. Although the thermodynamic limit inevitably implies an infinitely large Hilbert space, the latter limit exploits the intrinsically infinite  Hilbert space of one of its components, typically, a bosonic mode. 
Among the different systems exhibiting a finite-component QPT, we find the paradigmatic quantum Rabi (QRM)~\cite{Hwang:15,Puebla:16} and Jaynes-Cummings models~\cite{Hwang:16}. Such finite-component systems exhibit QPTs of different universality classes~\cite{Hwang:15,Hwang:16,Peng:19,Zhu:20}, as well as dissipative phase transitions~\cite{Hwang:18}. Beyond its fundamental relevance, this new route towards criticality can serve as an ideal testbed for theoretical and experimental investigations of QPTs, such as studies on their advantage for metrological purposes~\cite{Garbe:20,Chu:20} or the emergence of distinctive scaling laws in dissipative and critical dynamics~\cite{Puebla:20a}. Moreover, owing to the universality of phase transitions,  these finite-component systems can also be used to explore universal aspects of QPTs that will appear in the conventional thermodynamic limit of distinct systems~\cite{Puebla:17,PueblaThesis}.

Besides the singular behavior taking place in the ground state as a consequence of a QPT, a new class of critical phenomena has been formulated in the nonequilibrium dynamics of the a system followed by a sudden quench of an external parameter, dubbed dynamical quantum phase transitions (DPTs)~\cite{Heyl:13,Sharma:16,Budich:16,Lang:18b,Homrighausen:17,Halimeh:17,Lang:18,Zunkovic:18,Kosior:18,Kosior:18b,Jafari:19b,Jafari:19,Halimeh:19,Trapin:20,Wu:20,Ding:20,Porta:20} (see also the review~\cite{Heyl:18} and references therein). As for QPTs, in DPTs relevant quantities may exhibit a non-analytical behavior, yet they do so in the time domain~\cite{Heyl:18,Garrahan:10,Lesanovsky:13}. In particular, the elegant connection between the Loschmidt echo of a quenched state and the partition function with a complex temperature has allowed for the formulation of universality and scaling relations in DPTs~\cite{Heyl:15}. 
This new phenomenon of quantum matter out of equilibrium has been recently observed~\cite{Flaschner:18,Jurcevic:17,Zhang:17b,Tian:19,Wang:19,Guo:19}. Nevertheless, as in the context of QPTs, strictly non-analytical behavior is only expected in the thermodynamic limit.

In this article we show that DPTs can appear also in finite-component systems. That is,  DPTs can take place without scaling up the number of components, by only tuning the system parameters, and thus without altering the Hilbert space dimension. In particular, we demonstrate that all the phenomenology of DPTs applies to the QRM,  supporting our findings with numerical simulations and a semiclassical analysis.

{\em Dynamical quantum phase transitions.--} Similar to the phenomenology of standard QPTs, critical behavior can be found in the time domain, as for example in the nonequilibrium dynamics resulting from a sudden quench~\cite{Dorner:12,Heyl:13,Sharma:16,Budich:16,Lang:18,Heyl:18}. Typically, DPTs are studied by preparing the ground state $|\varphi_0(g_1)\rangle$ of a system at a certain value of an external parameter $g_1$, and then suddenly changing its magnitude to $g_2$. Note that DPTs have also been identified at non-zero temperature~\cite{Mera:18,Lang:18b}. The resulting dynamics may reveal the existence of two different types of DPTs, whose relation has been established in spin models~\cite{Homrighausen:17,Halimeh:17,Zunkovic:18}.

First, and in analogy with equilibrium QPTs, the different dynamical phases can be characterized by an order parameter, however, such quantity is defined as the long-time average of a certain observable, i.e.
\begin{align}\label{eq:Aover}
\overline{\langle \mathcal{A}\rangle} = \lim_{T\rightarrow \infty} \frac{1}{T}\int_0^T dt \langle \mathcal{A}\rangle_t,
  \end{align}
where $\langle\mathcal{A}\rangle_t$ refers to the expectation value of the observable $\mathcal{A}$ after an evolution time $t$. In the thermodynamic limit, denoted here by $\eta\rightarrow\infty$, this type of dynamical phase transition (DPT-I) takes place at $g_{2,c}$ such that $\overline{\langle \mathcal{A}\rangle}\neq 0$ for $g_2>g_{2,c}$ and zero otherwise. 

Second, dynamical critical behavior may also appear as non-analyticities of the Loschmidt echo rate function $\lim_{\eta\rightarrow \infty}r(t)$ at certain critical times $t_c$~\cite{Heyl:18}.  We refer here to this type of phase transition as DPT-II. The Loschmidt echo and its associated rate are defined as
\begin{align}\label{eq:rt}
L(t)=\langle \varphi_0(g_1)| e^{-it H(g_2)}|\varphi_0(g_1)\rangle,\quad r(t)=-\frac{1}{\eta}\log |L(t)|^{2},
  \end{align}
where we assume the ground state $\ket{\varphi_0(g_1)}$ of $H(g_1)$ to be evolving under $H(g_2)$.  The resemblance of $L(t)$ with a thermal partition function has motivated the connection between equilibrium and nonequilibrium phase transitions, which can be formalized by extending the time to the complex plane and considering boundary partition functions~\cite{Heyl:18,LeClair:95}. 

In a standard scenario, the variable $\eta$ refers to the number of constituents, so that a DPT takes place in the conventional thermodynamic limit $\eta\rightarrow\infty$~\cite{Heyl:18}. As anticipated, here we show that DPTs can appear in a system made of two degrees of freedom and where $\eta$ is associated with a ratio of frequencies appearing in the Hamiltonian rather than with the {\em size} of the system.


\begin{figure}[t!]
\centering
\includegraphics[width=0.75\linewidth]{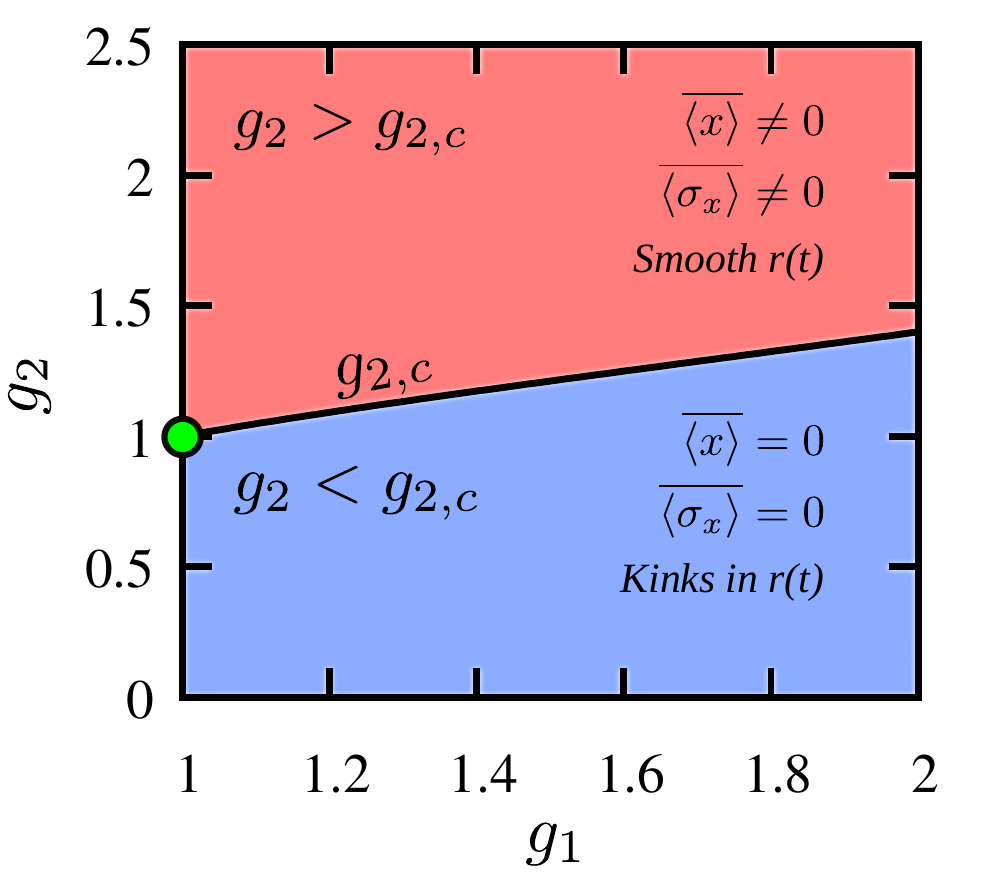}
\caption{\small{Schematic illustration of the dynamical phase diagram of the QRM. The symmetry-broken ground state of $H(g_1)$ at $g_1>1$ is quenched to $g_2$, which leads to the two types of DPTs, whose critical point takes place at $g_{2,c}=g_1(3+g_1^2)/(2(1+g_1^2))$: For $g_2>g_{2,c}$ the long-time averaged order parameters are non-zero, $\overline{\langle \sigma_x\rangle}\neq 0$ and $\overline{\langle x\rangle}\neq 0$, while they vanish for $g_{2}<g_{2,c}$ (DPT-I). In addition, the DPT-II is revealed in the rate function $r(t)$, which shows either a smooth behavior ($g_{2}>g_{2,c}$) or kinks and thus non-analytical ($g_{2}<g_{2,c}$). The equilibrium QPT at $g=1$ is indicated by a solid circle. See main text for further details. }}
\label{fig1}
\end{figure}

{\em Quantum Rabi model.--}  The QRM  describes the fundamental and ubiquitous interaction of a spin with a single bosonic mode~\cite{Rabi:36,Rabi:37}, whose Hamiltonian can be written as
\begin{equation}
\label{eq:HRabi}
H(g)=\frac{\Omega}{2}\sigma_z+\omega_0\adaga-g\frac{\sqrt{\Omega\omega_0}}{2}\left(a+\adag \right)\sigma_x,
\end{equation}
being $\Omega$ and $\omega_0$ the corresponding frequencies of the spin and single mode, respectively, and with $g$  a dimensionless coupling strength. The spin is described by the standard Pauli matrices,  $\sigma_{\alpha}$ with $\alpha\in\{x,y,z\}$ so that $\sigma_z=\ket{\uparrow}\bra{\uparrow}-\ket{\downarrow}\bra{\downarrow}$, while the operators $a$ and $\adag$, such that $[a,\adag]=1$, refer to the bosonic mode.


Despite counting only with two degrees of freedom, the QRM shows a QPT at the critical value $g=1$, as demonstrated in~\cite{Hwang:15}. Indeed, in the limit  $\eta\equiv \Omega/\omega_0\rightarrow \infty$, the $H(g)$ exhibits two distinct phases. In the superradiant phase, for $g>1$, the bosonic mode acquires coherence $\langle a\rangle \propto \sqrt{\eta}$ and the $Z_2$ parity symmetry is spontaneously broken, which also extends to excited states~\cite{Puebla:16}. The exact solution of the QRM in the limit $\eta\rightarrow\infty$ allows us to obtain the exact symmetry-breaking ground states for $g>g_c=1$, which read as \begin{align}\label{eq:GSmt}
  |\varphi_0^{\pm}(g)\rangle=\mathcal{D}[\pm \alpha_{\rm sp}(g)]\mathcal{S}[s_{\rm sp}(g)]\ket{0}\ket{\downarrow^{\pm}},
\end{align} where $\mathcal{D}[\alpha]=e^{\alpha a^\dagger-\alpha^*a}$ and $\mathcal{S}[s]=e^{({s^*a^{\dagger,2}-sa^2})/2}$ are the displacement and squeezing operators, respectively, and $\ket{\downarrow^{\pm}}=\pm\sqrt{{1-g^{-2}}/2}\ket{\uparrow}+\sqrt{(1+g^{-2})/2}\ket{\downarrow}$~\cite{Hwang:15,sup}. The amplitudes are $\alpha_{\rm sp}(g)=\sqrt{\eta}\sqrt{(g^2-g^{-2})/4}$ and $s_{\rm sp}(g)=-1/4\log(1-g^{-4})$. In the reminder of this article, we will constrain ourselves to the positive choice in Eq.~\eqref{eq:GSmt}, since $|\varphi^-_0(g)\rangle$ leads to completely equivalent results.

In order to investigate the DPTs, we  proceed as follows. First, the symmetry-broken ground state $|\varphi^{+}_0(g_1)\rangle$ with $g_1>1$ is prepared. Second, the state is  quenched to $g_2$, i.e., letting evolve $|\varphi^+_0(g_1)\rangle$ under $H(g_2)$. Depending on the chosen pair of values $g_1$ and $g_2$, the dynamics will be in a different dynamical phase. As we show in the following, the resulting dynamical phase diagram of the QRM is illustrated in Fig.~\ref{fig1}. There exists a critical coupling $g_{2,c}$ which divides the phase diagram, and depends on $g_1$. For $g_2>g_{2,c}$ the long-time averaged order parameters are non-zero and the rate function is smooth, while for $g_2<g_{2,c}$ the order parameters vanish and $r(t)$ becomes non-analytical at certain critical times.

Such critical line $g_{2,c}$ can be derived from the semiclassical structure of the QRM in the $\eta\rightarrow\infty$ limit~\cite{sup}. Indeed, the sudden quench of the initial state $|\varphi^+_0(g_1)\rangle$ to $g_2$ produces work onto the system, leading to a final energy $E(g_1,g_2)=\langle \varphi_0^{\pm}(g_1)|H(g_2)|\varphi_0^{\pm}(g_1)\rangle$. The larger the quench the more energy is transferred into the system. Eventually, the quenched state acquires an energy $E(g_1,g_2)>E_c$ where $E_c=-\eta\omega_0/2$ denotes the energy of the separatrix in the double-well structure. For $E(g_1,g_2)<E_c$ the phase space consists of disconnected regions, while above $E_c$ these regions are merged into a single one~\cite{sup}. At the quantum level, the eigenstates of $H(g)$ cease to be two-fold degenerate~\cite{Puebla:16}, which is a signature of the excited-state QPT~\cite{Heiss:02,Leyvraz:05,Cejnar:06,Cejnar:08,Caprio:08,Puebla:13,Brandes:13}. From the condition $E(g_1,g_{2,c})=E_c$ one obtains at leading order in $\eta$~\cite{sup}
\begin{align}\label{eq:g2c}
g_{2,c}=\frac{g_1(3+g_1^2)}{2(1+g_1^2)}, \quad {\rm with} \quad g_1>1.
\end{align}
It is worth noting that the location of the DPTs, i.e. the dynamical critical coupling $g_{2,c}$, is always larger than the critical point of its equilibrium counterpart $g=1$ (cf. Fig.~\ref{fig1}). In the following we analyze the two aforementioned DPTs in the QRM.

{\em DPT-I.--} We start by considering the long-time averaged order parameters. In analogy with the equilibrium QPT of the QRM, we take $\sigma_x$ and $x=(a+\adag)/\sqrt{2\eta}$ whose ground-state expectation values serve as good order parameters of the symmetry-breaking QPT at $g=1$~\cite{Hwang:15}. We compute $\overline{\langle\sigma_x \rangle}$ and $\overline{\langle x\rangle}$ as in Eq.~\eqref{eq:Aover} from the sudden quench dynamics of  $|\varphi_0^+(g_1)\rangle$ evolving under $H(g_2)$. 

As soon as $g_2<g_{2,c}$, the state is quenched onto excited states which are not two-fold degenerate, and thus no longer support a non-zero order parameter. In particular, $e^{-itH(g_2)}|\varphi_0^+(g_1)\rangle=\sum_{n}e^{-it E_n(g_2)}\langle \phi_n|\varphi_0^+(g_1)\rangle |\phi_n\rangle$ with $H(g_2)=\sum_n E_n(g_2)\ket{\phi_n(g_2)}\bra{\phi_n(g_2)}$. Since $E(g_1,g_2)>E_c$, the populated eigenstates $E_n>E_c$ conserve the $Z_2$ parity symmetry so that $\bra{\phi_n(g_2)}\sigma_x\ket{\phi_n(g_2)}=0$, and similarly for $x$~\cite{Puebla:16}. As a consequence, their long-time averaged values vanish. To the contrary, when $g_{2}>g_{2,c}$, the quenched stated populates eigenstates of $H(g_2)$ that are two-fold degenerate as the ground state, and thus long-lived symmetry-breaking states can persist~\cite{Puebla:13}. 

In Fig.~\ref{fig2}(a) and (b) we show these long-time averaged order parameters, $\overline{\langle \sigma_x \rangle}$ and $\overline{\langle x \rangle}$, for a reasonably large frequency ratio, $\eta=100$. As an example, we chose $g_1=3/2$ which yields a critical value $g_{2,c}\approx 1.21$ (cf. Eq.~\eqref{eq:g2c}). The dramatic change in both order parameters is clearly visible at this point. The dynamics of the order parameters are also plotted in Fig.~\ref{fig2}(c) and (d), which illustrate their remarkable  different behavior depending on $g_2$. The solid line in Fig.~\ref{fig2}(a) and (b) corresponds to a semiclassical description, which is achieved in a standard manner (see~\cite{sup} for further details). The semiclassical curve agrees well with the fully quantum mechanical results. 

A closer inspection around the dynamical critical point $g_{2,c}$ reveals that the long-time averaged order parameters under the semiclassical approximation do not vanish in a continuous fashion but rather abruptly jump at $g_{2,c}$. As soon as the available phase space joins the disconnected regions, i.e. when $E(g_1,g_2)>E_c$, the long-time averaged order parameters become zero, while at $g_{2,c}$ we obtain $\overline{\langle \sigma_x \rangle}\neq 0$  and $\overline{\langle x \rangle}\neq 0$~\cite{sup}. Such a discontinuous transition is difficult to corroborate in the finite-$\eta$ quantum dynamics.
However, from standard finite-size  scaling theory of phase transitions~\cite{Fisher:74,Fisher:72} it follows that  if $\overline{\langle\mathcal{A}\rangle}\propto |g-g_c|^\gamma$  with $\gamma>0$ in the thermodynamic limit, then its finite-$\eta$ value at $g_c$ goes to zero as $\overline{\langle\mathcal{A}\rangle}\propto \eta^{-\nu/\gamma}$ with $\nu$ the correlation length critical exponent. In this case, by increasing $\eta$ the quantities  $\overline{\langle \sigma_x \rangle}$ and $\overline{\langle x \rangle}$ at $g_{2,c}$ do not show this behavior, but rather an increase to a constant non-zero value~\cite{sup}. This suggests that the QRM exhibits a first-order finite-component DPT-I at $g_{2,c}$. In addition, we comment that the signatures of the DPT-I are also visible in other quantities, not related with symmetry-breaking order parameters, such as $\sigma_z$ and $\adaga$~\cite{sup}.
%
%
%

\begin{figure}[t!]
\centering
\includegraphics[width=1.\linewidth]{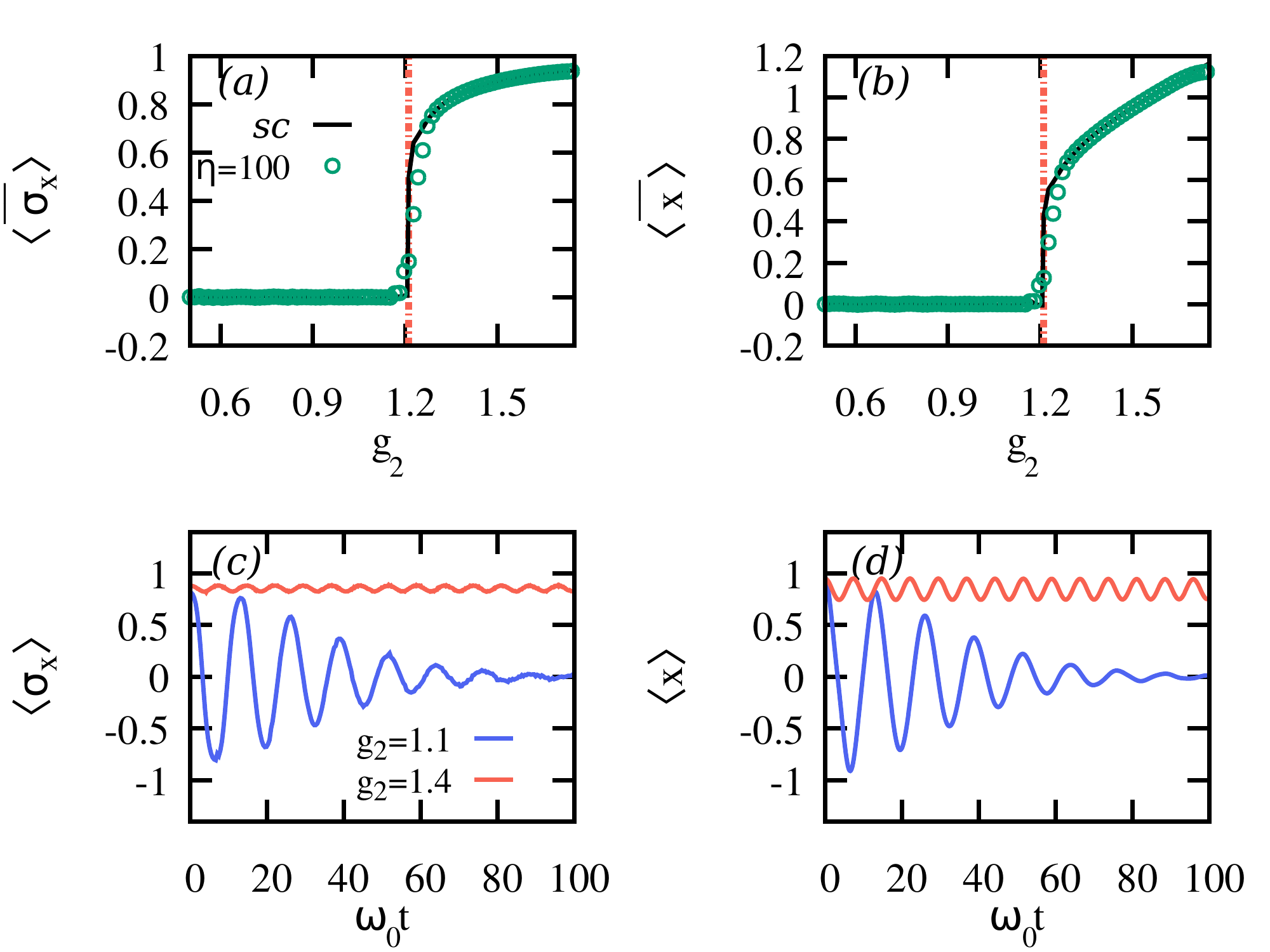}
\caption{\small{Panels (a) and (b) show the order parameter of the DPT-I, $\overline{\langle \sigma_x\rangle}$ and $\overline{\langle x\rangle}$, respectively, as a function of the quenched coupling $g_2$. The ground state corresponds to $g_1=3/2$, so that $g_{2,c}\approx 1.21$ (dashed red line). The solid points correspond to long-time averaged for $\eta=100$, obtained in the time window $\omega_0 t\in[100,500]$. The solid line has been obtained using a semiclassical approximation. Panels (c) and (d) show the actual quantum dynamics for $\langle \sigma_x\rangle$ and $\langle x\rangle$  at either side of the critical point, $g_2<g_{2,c}$ (blue lines) and $g_2>g_{2,c}$ (red lines). See main text for further details. }}
\label{fig2}
\end{figure}

{\em DPT-II.--} Let us now analyze the critical behavior appearing in the Loschmidt echo. As done previously, we quench the initial state $|\varphi_0^+(g_1)\rangle$ under $H(g_2)$ but now turn our attention to the rate function $r(t)$. As the  ground state is two-fold degenerate, the overlap $|L(t)|^2$ adopts the form~\cite{Heyl:14,Zunkovic:18}
\begin{align}
|L(t)|^2=\sum_{q=+,-}|\langle \varphi_0^{q}(g_1)|e^{-it H(g_2)}|\varphi_0^+(g_1)\rangle|^2.
\end{align}
In the limit $\eta\rightarrow \infty$, the rate $r(t)$ becomes non-analytical at certain critical times $t_c$ as long as $g_2<g_{2,c}$. 
In Fig.~\ref{fig3}(a) we show an example with $\eta=100$ and $g_1=3/2$ and representative of the dynamics for the two phases, $g_2=0.75$ and $g_2=1.4$ since $g_{2,c}\approx 1.21$. For $g_2<g_{2,c}$ the rate $r(t)$ displays {\em kinks} at different times,  while it is smooth otherwise.

\begin{figure}[t!]
\centering
\includegraphics[width=1.\linewidth]{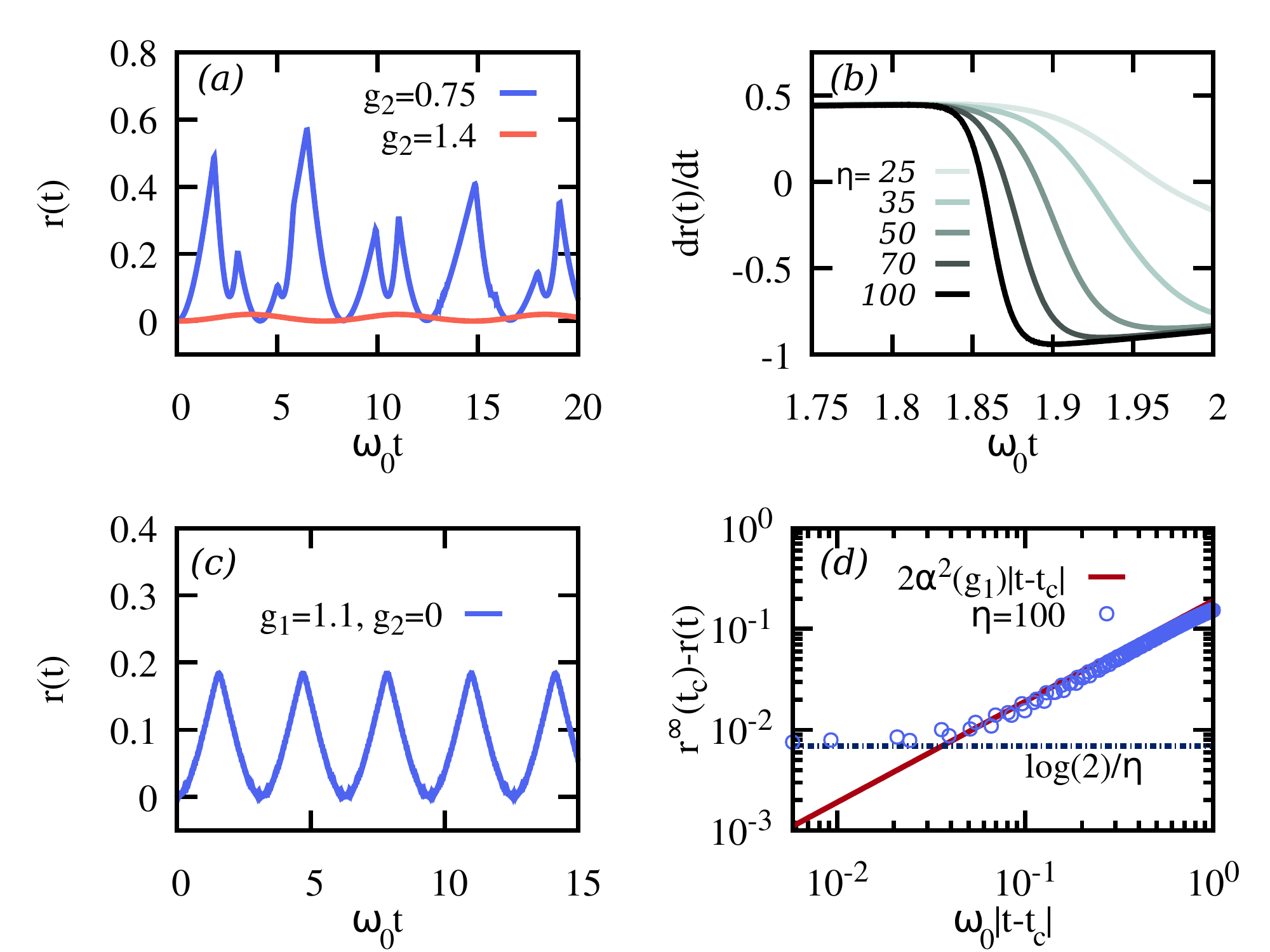}
\caption{\small{(a) Rate $r(t)$ for $\eta=100$ and $g_1=3/2$ and two different final coupling strength $g_2$. For $g_2>g_{2,c}$ (red) the rate function is a smooth and periodic function, while for $g_2<g_{2,c}$ (blue) the rate $r(t)$ exhibits kinks at certain critical times. The larger $\eta$ the sharper the kinks. This is illustrated in panel (b), where the slope $dr(t)/dt$ for $g_2=0.75$ and for various $\eta$ values is plotted as a function of $\omega_0t$ close to the first kink in $r(t)$ whose critical time is $\omega_0t_c\approx 1.85$. Panel (c) shows the rate for $\eta=100$, $g_2=0$ and $g_1=1.1$. The critical times $\omega_0 t_c=\pi/2(2n+1)$ are visible, which have an associated finite-$\eta$ scaling and critical exponent, as shown in (d) for $\omega_0 t_c=\pi/2$. The horizontal dashed line corresponds to the finite-$\eta$ value, while the solid line shows the expected scaling in the limit $\eta\rightarrow \infty$.  See main text for further details.}}
\label{fig3}
\end{figure}

These results can be understood as follows. As soon as $g_2<g_{2,c}$, the state acquires sufficient energy so that it can explore the whole phase space. Thus, the quenched state will exhibit a non-zero overlap with either of the ground states but at different times. The states $|\varphi_0^{\pm}\rangle$ comprise a displaced bosonic mode by an amount $|\alpha|\propto \sqrt{\eta}$, so that  $P_{\pm}(t)=|\langle \varphi_0^{\pm}(g_1)|e^{-itH(g_2)}|\varphi_0^+(g_1)\rangle|^2\propto e^{-\eta f_{\pm}(t)}$ with an $\eta$-intensive function $f_{\pm}(t)$. In the limit $\eta\rightarrow\infty$ only one of them dominates, and thus $r^{\infty}(t)\equiv \lim_{\eta\rightarrow\infty}r(t)={\min_{q=\pm}}f_{q}(t)$. Critical times in the rate function appear therefore when $f_{+}(t_c)=f_-(t_c)$. To the contrary, for $g_{2}>g_{2,c}$, the quenched state is locked within a symmetry-broken phase, or in semiclassical terms, within one of the two disjoint phase-space islands. Hence, in our case the rate is only dominated by $P_+(t)$, and thus $r(t)$ results simply in a smooth and periodic function.

Similar to finite-size effects in traditional many-body systems, a finite value of $\eta$ introduces corrections to the strictly non-analytical behavior. In order to illustrate this  effect on the rate function, we compute $dr(t)/dt$ close to the first kink $\omega_0t_c\approx 1.85$ in Fig.~\ref{fig3}(a). This is plotted in Fig.~\ref{fig3}(b). As shown above, in the limit $\eta\rightarrow\infty$ the rate becomes non-analytical and $dr(t)/dt$ will show a discontinuity at $t_c$. The onset of this discontinuity is clearly visible in Fig.~\ref{fig3}(b), which shows how $dr(t)/dt$ becomes increasingly sharper around $\omega_0t_c$ for increasing $\eta$, from $25$ to $100$.

The case $g_2=0$ deserves special mention due to its simplicity. In this case, we find $f_{\pm}(t)=(g_1^2-g_1^{-2})(1\pm \cos\omega_0t)/2$, and thus  $r(t)$ becomes non-analytical at the critical times $\omega_0 t_c=\pi/2+n\pi$ for $n=0,1,\ldots$, for any $g_1>1$, whose value amounts to $r^\infty(t_c)=(g_1^2-g_1^{-2})/2$~\cite{sup}. In this case we can also obtain the finite-$\eta$ corrections, which modify the critical rate as $r(t_c)=r^{\infty}(t_c)-\log2/\eta$. Moreover, a straightforward calculation allows us to obtain the critical exponent $\beta=1$, such that $r(t_c)-r(t)\sim |t-t_c|^{\beta}$ for $|t-t_c|\ll 1$. This is plotted in Fig.~\ref{fig3}(c) and (d), while we refer the interested readers to~\cite{sup} for further details.

Finally, it is worth stressing that both DPTs as described here are washed out when suppressing the counter-rotating terms in Eq.~\eqref{eq:HRabi}, i.e., when reducing the QRM to a Jaynes-Cummings model~\cite{Jaynes:63}. Indeed, the reported DPTs in the QRM appear in the ultra- and deep-strong coupling regimes, while their absence in the Jaynes-Cummings stems from the additional symmetry of this model ($U(1)$), which prevents the quenched state to access subspaces with distinct order parameter~\cite{Heyl:18}. Recall that the Jaynes-Cummings model does display an {\em equilibrium} finite-component QPT~\cite{Hwang:16}. This suggests that finite-component DPTs must be inspected in a case-by-case basis, as they do not necessarily accompany its equilibrium QPT counterpart, just like standard DPTs~\cite{Heyl:18}.

{\em Conclusions.--} We have shown that dynamical quantum phase transitions can be studied and realized in systems comprising only few and finite constituents, where critical behavior is attained without modifying the Hilbert space dimension or scaling up the number of components. In particular, by inspecting the quantum Rabi model which only involves two subsystems, namely, a spin and a bosonic mode, we demonstrate the presence of the two types of dynamical quantum phase transitions. This takes place in the same parameter limit in which the quantum Rabi model features a quantum phase transition, i.e.  singular ground and excited states~\cite{Hwang:15,Puebla:16}. Depending on the strength of a sudden quench, the resulting nonequilibrium dynamics fall into either of the two dynamical phases, that is, either in a phase with i) non-zero long-time averaged order parameters and a smooth rate function, or in the other phase with ii) vanishing long-time averaged order parameters and \emph{kinks} or non-analytic behavior emerging in the rate function. A semiclassical analysis reveals an abrupt jump of the long-time averaged order parameters, while a critical exponent and corrections to the infinite limit are derived for the non-analytical rate function.

Owing to the relevance of the quantum Rabi model in a variety of platforms~\cite{Crespi:12,Langford:17,Lv:18}, the reported results are amenable for their experimental realization. Our findings open therefore new avenues for the exploration of dynamical quantum phase transitions without the need to scale up the number of components, thus allowing for the investigation of these critical phenomena and the onset thereof in well controllable systems.

\begin{acknowledgments}
R. P. thanks Mauro Paternostro for useful comments, and also acknowledges the support by the SFI-DfE Investigator Programme (grant 15/IA/2864).
\end{acknowledgments}

%


\newpage

\setcounter{equation}{0}
\setcounter{figure}{0}
\setcounter{table}{0}
\makeatletter
\renewcommand{\theequation}{S\arabic{equation}}
\renewcommand{\thefigure}{S\arabic{figure}}
\renewcommand{\bibnumfmt}[1]{[S#1]}
\renewcommand{\citenumfont}[1]{S#1}

\begin{widetext}
  \section{Supplemental Material \\ Finite-Component Dynamical Quantum Phase Transitions}
  \begin{center}
    Ricardo Puebla\\ \vspace{0.1cm}
    {\em {\small Centre for Theoretical Atomic, Molecular and Optical Physics, School of Mathematics and Physics,\\ Queen's University Belfast, Belfast BT7 1NN, United Kingdom}}\\
    \end{center}
  \maketitle

  \section{I. Effective model and ground states of the Quantum Rabi model}
In the limit $\eta\equiv \Omega/\omega_0\rightarrow \infty$ a Schrieffer-Wolff transformation of the QRM, as given in Eq. (3) of the main text, allows us to find an effective low-energy model that reveals a QPT at $g=1$. For that, we require $[H_0,S]=\lambda \sigma_x(a+\adag)$ with $H_0=\Omega/2\sigma_z+\omega_0\adaga$, so that upon $e^{-S}He^S$ the Hamiltonian is diagonal at leading order in $\eta^{-1}$ (see~\cite{Hwang:15} for further details). In particular, we find $S=ig/2\eta^{-1/2}\sigma_y(a+\adag)$ which leads to
\begin{align}
H_{\rm np}=\frac{\Omega}{2}\sigma_z+\omega_0\adaga+\frac{g^2\omega_0}{4}\sigma_z(a+\adag)^2,
  \end{align}
valid for $0\leq g\leq 1$. For $g>g_c=1$, one needs first to displace the bosonic mode and rotate the spin degree of freedom to then apply again the Schrieffer-Wolff transformation. Following this procedure, one finds the displacement amplitude $\alpha_{\rm sp}(g)=\sqrt{\eta}\sqrt{(g^2-g^{-2})/4}$, while the spin is rotated according to $|\downarrow^{\pm}\rangle=\pm \sqrt{(1-g^{-2})/2}|\uparrow\rangle+\sqrt{(1+g^{-2})/2} |\downarrow\rangle$. The two-equivalently displaced effective Hamiltonians in the low-energy subspace read as
\begin{align}
H_{\rm sp}^{\pm}=\mathcal{D}^\dagger[\pm \alpha_{\rm sp}(g)]H\mathcal{D}[\pm\alpha_{\rm sp}(g)]=\omega_0\adaga+\frac{\omega_0}{4g^4} \tilde{\sigma}_z^{\pm}(a+\adag)^2-\frac{\Omega}{4}(g^2+g^{-2}),
  \end{align}
with $\tilde{\sigma}_z^{\pm}=\ket{\uparrow^{\pm}}\bra{\uparrow^{\pm}}-\ket{\downarrow^{\pm}}\bra{\downarrow^{\pm}}$ and $\mathcal{D}[\alpha]=e^{\alpha \adag -\alpha^*a}$ the displacement operator. Both effective Hamiltonians are diagonal now in the spin basis, which upon projecting onto the low-energy state, the Hamiltonian is quadratic in $a$ and $\adag$, and thus easily diagonalizable. The ground state of the QRM in the $\eta\rightarrow\infty$ limit is given by
\begin{align}\label{eq:GSSB}
  \ket{\varphi_0(g)}=\begin{cases}\mathcal{S}[s_{\rm np}(g)]\ket{0}\ket{\downarrow},\qquad \qquad \quad 0\leq g \leq 1\\
  \mathcal{D}[\pm \alpha_{\rm sp}(g)]\mathcal{S}[s_{\rm sp}(g)]\ket{0}\ket{\downarrow^{\pm}},\ \quad g>1,
  \end{cases}
\end{align}
with  $\mathcal{S}[s]=e^{s/2(a^{\dagger,2}-a^2)}$ the squeezing operator, for real amplitude $s$, and $s_{\rm np}(g)=-1/4\log(1-g^2)$, and $s_{\rm sp}(g)=-1/4\log(1-g^{-4})$~\cite{Hwang:15}. For $g>1$, $|\varphi_0^{\pm}(g)\rangle$ denotes the two symmetry-breaking states.

The initial symmetry-breaking quantities are therefore (starting with the ground state of $H(g)$ as given in Eq.~\eqref{eq:GSSB})
\begin{align}
  \langle \varphi_0(g)|\sigma_x|\varphi_0(g)\rangle&=\begin{cases} 0, \qquad \qquad \qquad \quad 0\leq g \leq 1\\ \pm \sqrt{1-g^{-4}} ,\ \qquad \ \ \, \quad g>1.
  \end{cases}\\
  \langle \varphi_0(g)|(a+\adag)|\varphi_0(g)\rangle&=\begin{cases} 0, \qquad \qquad \qquad \quad 0\leq g \leq 1\\ \pm \sqrt{\eta}\sqrt{g^2-g^{-2}},\ \ \ \ \quad g>1.
  \end{cases}
  \end{align}
As commented in the main text, we initialize the system in the ground state $|\varphi_0^{+}(g>1)\rangle$.

\section{II. Dynamical Phase Diagram of the Quantum Rabi Model}
The energy of the quenched initial state $|\varphi_0(g_1)\rangle$ is given by
\begin{align}
  E(g_1,g_2)=\langle \varphi_0(g_1)|H(g_2)|\varphi_0(g_1)\rangle=\begin{cases} -\frac{\eta\omega_0}{2}+\omega_0\sinh^2(s_{\rm np}(g_1)),\qquad g_1\leq 1\\
  -\frac{\eta\omega_0}{2g_1^2}+\omega_0 (\sinh^2(s_{\rm sp}(g_1))+\alpha^2_{\rm sp}(g_1))-\frac{g_2\omega_0\eta}{2}\ g_1\left(1-\frac{1}{g_1^4}\right), \qquad g_1>1
\end{cases}
  \end{align}
 Substituting the expressions for $s_{\rm sp}(g)$ and $\alpha_{\rm sp}(g)$, and assuming $\eta\gg 1$, we obtain that a quench state $|\varphi_0^{\pm}(g_1>1)\rangle$ acquires an energy $E(g_1,g_{2,c})= -\eta \omega_0/2$ (at leading order in $\eta$) for a critical quench parameter $g_{2,c}$
\begin{align}
g_{2,c}=\frac{g_1(3+g_1^2)}{2(1+g_1^2)},
\end{align}
with $g_1> 1$. This is the expression given in the main text. The energy $E=-\eta\omega_0/2$ corresponds to the separatrix in the phase space (cf. Fig.~\ref{figSM1}). Note that we have neglected the contribution $\omega_0\sinh^2(s_{\rm sp}(g_1))$ to the total energy, which diverges for $g_1=1$. Yet, for any $g_1\neq 1$, the limit $\eta\rightarrow \infty$ dominates and the previous analysis is valid.  In this manner, $E(g_1,g_2>g_{2,c})<-\eta\omega_0/2$ and there is no change in dynamical phase phase (dynamical order parameter different from zero), while for $E(g_1,g_2<g_{2,c})>-\eta\omega_0/2$ the dynamical order parameter becomes zero. This is further illustrated by means of a semiclassical approximation.

\begin{figure}[t!]
\centering
\includegraphics[width=1\linewidth]{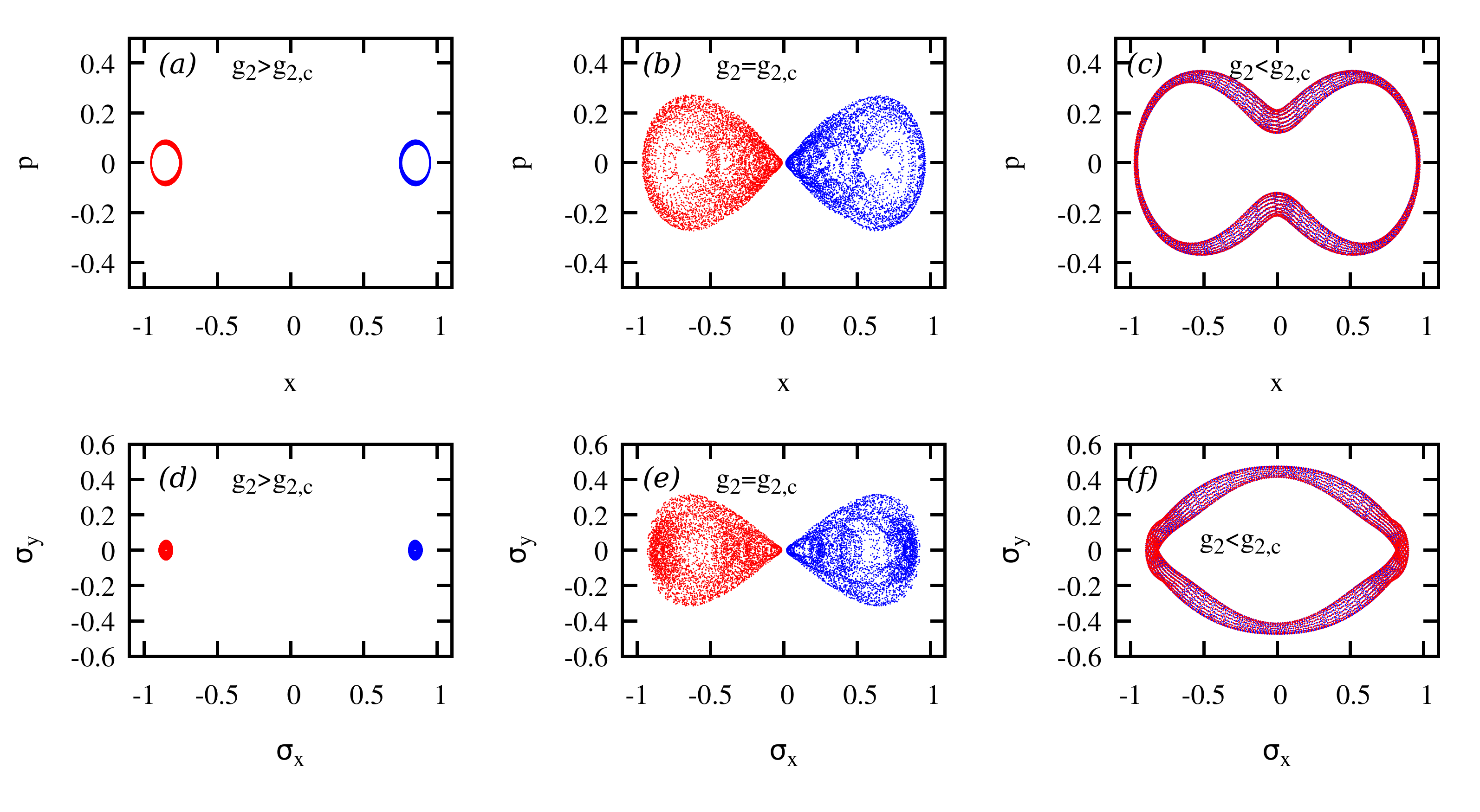}
\caption{\small{Snapshots or Poincar{\'e} sections of the semiclassical quench dynamics for $(x,p)$ (top panels) and $(\sigma_x,\sigma_y)$ (bottom panels), obtained solving Eqs.~\eqref{eq:xdot}-\eqref{eq:szdot} and taking $5000$ points up to $\omega_0t_f=10^3$. The value $\sigma_z$ follows from $\sigma_x^2+\sigma_y^2+\sigma_z^2=1$. Left, middle and right columns show the Poincar{\'e} sections for different quench values $g_2$, namely, $g_2>g_{2,c}$, $g_2=g_{2,c}$ and $g_2<g_{2,c}$, respectively. Red and blue dots correspond to the initial condition, $x>0$ and $\sigma_x>0$ (blue) and $x<0$ and $\sigma_x<0$ (red). }}
\label{figSM1}
\end{figure}

\section{III. Semiclassical approximation and Poincar{\'e} sections}
The semiclassical equations of motion for the QRM in the $\eta\rightarrow\infty$ are
\begin{align}\label{eq:xdot}
  \dot{x}(t)&=p(t)\\ \dot{p}(t)&=-x(t)+g\frac{\sigma_x(t)}{\sqrt{2}}\\
  \dot{\sigma}_x(t)&=-\sigma_y(t)\\
  \dot{\sigma}_y(t)&=\sigma_x(t)+g\sqrt{2}x(t)\sigma_z(t)\\
  \dot{\sigma}_z(t)&=-\sqrt{2}g x(t)\sigma_y(t).\label{eq:szdot}
  \end{align}
where $(x,p)=\eta^{-1/2}(\tilde{x},\tilde{p})$ are the rescaled and semiclassical continuous variables of the harmonic oscillator $\tilde{x}=(a+\adag)/\sqrt{2}$ and $\tilde{p}=i(\adag-a)/\sqrt{2}$, so that $[x,p]=0$. The rescaled Hamiltonian reads therefore as $H=(x^2+p^2)/2 +\sigma_z/2-gx\sigma_x/\sqrt{2}$, so that $dH/dt=0$. As a result of the semiclassical treatment, the spin degree of freedom is completely factorized and thus, $\sum_{\alpha=x,y,z}\sigma_\alpha^2(t)=1$.

Setting the initial condition as given in Sec. I, we solve the dynamics dictated by Eqs.~\eqref{eq:xdot}-\eqref{eq:szdot}. The long-time averaged values can be then computed as for the quantum case (cf. Fig. 2(a) and (b) of the main text).  It is instructive to compute  the Poincar{\'e} sections obtained in the semiclassical quench dynamics. These sections are plotted in Fig.~\ref{figSM1} for the two different initial symmetry-breaking conditions, and distinct quench value $g_2$. The separatrix at $g_2=g_{2,c}$ is perfectly visible. As soon as $g_2<g_{2,c}$, the previously disconnected regions (cf. Fig.~\ref{figSM1}(a) and (b)) are merged into a single one (cf. Fig.~\ref{figSM1}(c) and (f)).  As a consequence the long-time averaged order parameters vanish if $g_{2}<g_{2,c}$, while they remain at a non-zero value if $g_{2}>g_{2,c}$.

\begin{figure}[t!]
\centering
\includegraphics[width=1\linewidth]{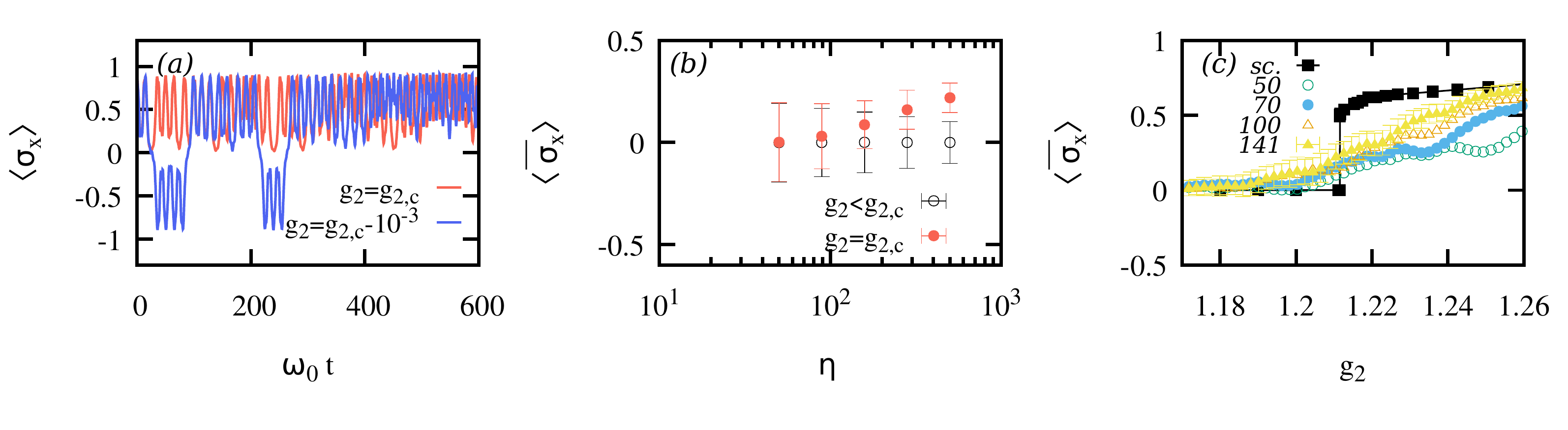}
\caption{\small{(a) Semiclassical evolution of $\sigma_x$, obtained from Eqs.~\eqref{eq:xdot}-\eqref{eq:szdot}. At the critical point $g_{2}=g_{2,c}$ the trajectory is still confined in one region (cf. Fig.~\ref{figSM1}), while a just a small variation $g_{2,c}-10^{-3}$ allows the system to explore for regions, resulting in jumps in the evolution. In panel (b) we show the long-time averaged value of $\langle \sigma_x\rangle$ for the quantum dynamics as a function of the $\eta$ value. For $g_{2,c}<g_{2,c}$, $\overline{\langle\sigma_x \rangle}$ remains at zero, while at $g_{2,c}$ the value increases for larger  $\eta$, suggesting a saturation value as in a discontinuous phase transition (here $g_1=4/3$). Panel (c) shows a zoom of $\overline{\langle\sigma_x \rangle}$ close to the transition point, $g_{2,c}\approx 1.21$ for various $\eta$ values and the semiclassical approximation showing the jump at $g_{2,c}$. The average is taken in the time window $\omega_0t\in [100,500]$ taking $400$ points. For a better illustration we display error bars only for the case with $\eta=141$.}}
\label{figSM2}
\end{figure}

\begin{figure}[t!]
\centering
\includegraphics[width=0.75\linewidth]{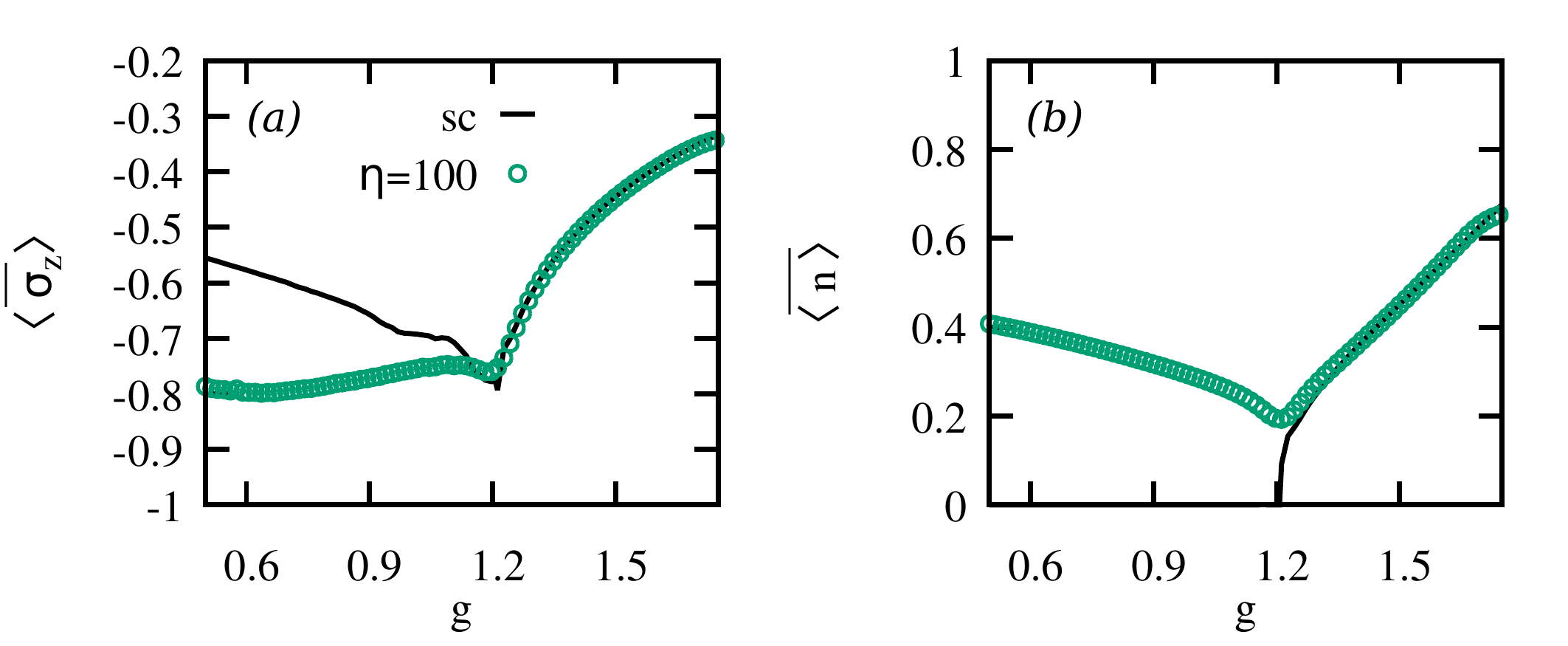}
\caption{\small{(a) Long-time averaged $\sigma_z$ for the quantum dynamics for a QRM with $\eta=100$ (points) and its semiclassical description (solid line), choosing $g_1=3/2$ and different $g_2$ values (same parameters as in Fig. 2 of the main text). The dynamical critical point $g_{2,c}\approx 1.21$ appears as a dip in the long-time average value. The disagreement between the semiclassical and the quantum results suggests that the quantum correlations are essential to properly describe this quantity in this dynamical phase, $g_2<g_{2,c}$. In panel (a) we show the long-time averaged value of the number of bosonic excitations $\overline{\langle n\rangle}$. Again, for $g_2>g_{2,c}$ the agreement is very good, while it fails for $g_2<g_{2,c}$.}}
\label{figSM3}
\end{figure}

\section{IV. Discontinuous long-time averaged order parameters and further signatures}
The ensemble average of a quantity $\mathcal{A}$ in the semiclassical approximation reads as
\begin{align}
\langle \mathcal{A}\rangle = \frac{\int d{\bf \xi} \rho[\xi,t]\mathcal{A}(\xi,t)}{\int d{\bf \xi}\rho[\xi,t]}
\end{align}
where $\rho[\xi,t]$ denotes the phase-space density evolving in time and ${\bf \xi}$ the phase-space variables, ${\bf \xi}=(x,p,\sigma_x,\sigma_y,\sigma_z)$.  By virtue of the Liouville's theorem since we quench an initial state following Hamiltonian dynamics, the denominator is constant. From the previous considerations, it follows that the semiclassical long-time averaged order parameters for these system with two degrees of freedom display an abrupt jump at $g_{2,c}$. For $g_{2}\geq g_{2,c}$ the phase-space integral must restricted to one of the disconnected islands (cf. Fig.~\ref{figSM1}), which entails a non-zero value. For $g_2<g_{2,c}$ both regions become connected so that $\overline{\langle \sigma_x\rangle}$ and $\overline{\langle x\rangle}$ vanish. See Fig.~\ref{figSM2}(a) where the semiclassical trajectory for $\sigma_x$ is plotted. For $g_{2}\lesssim g_{2,c}$ the dynamics displays jumps between the two regions, while for $g_{2}\geq g_{2,c}$ the order parameter $\sigma_x$ is always positive (or negative depending on the initial condition).

The quantum analogue of a discontinuous DPT-1 is harder to corroborate. On top of the required long-time average, the quantum case is prone to finite-$\eta$ corrections (cf. Fig.~\ref{figSM2}(b) and (c)).
 We remark that, although a Schrieffer-Wolff transformation allows us to analytically describe the low-energy subspace of the QRM in the limit $\eta\rightarrow \infty$, the quench dynamics involve high-energy states. An analytical treatment of the quench dynamics would require therefore to solve the dynamics of the full QRM, not only in the low-energy subspace. However, a finite-$\eta$ analysis suggests that $\overline{\langle \sigma_x\rangle}$ is non-zero at $g_{2,c}$ for increasing $\eta$, while for $g_{2}<g_{2,c}$ the value $\overline{\langle \sigma_x\rangle}$ is zero. Should the transition be continuous, the finite-$\eta$ would show a power-law behavior such that $\overline{\langle \sigma_x\rangle}\rightarrow 0$ for $\eta\rightarrow \infty$ (see main text), suggesting that the long-time averaged order parameters will display a  discontinuous jump in the limit $\eta\rightarrow \infty$, in the same manner as the semiclassical approximation. Yet, a more thorough investigation of the order of the DPT-1 (whether first-order or continuous) is left for future investigations.

 As commented in the main text, the signatures of the DPT-1 is also visible in quantities that are not related to symmetry breaking. This is indeed the case for $\sigma_z$ and $n\equiv \adaga$, which are plotted in Fig.~\ref{figSM3}. The DPT-1 appears as a dip in these quantities. Note that the semiclassical description fails to reproduce the quantum behavior for  $g_{2}<g_{2,c}$, which suggests that quantum correlations are essential to properly account for these quantities in this dynamical phase.

\section{V. Non-analytical rate function}



The rate function $r(t)$ of the Loschmidt echo is obtained from
\begin{align}
  |L(t)|^2&=\sum_{q=+,-}|\langle \varphi_0^{q}(g_1)|e^{-it H(g_2)}|\varphi_0^+(g_1)\rangle|^2.
\end{align}
as $r(t)=-1/\eta \log |L(t)|^2$. By assuming $g_1>1$ so that the squeezing parameter $s_{\rm sp}(g_1)\ll 1$, we can approximate the ground state $|\varphi_0^{\pm}(g_1)\rangle\approx |\pm\alpha_{\rm sp}(g_1)\rangle |\downarrow^{\pm}\rangle$. For $g_2=0$ the previous expression reduces to
\begin{align}\label{eq:Lta}
  |L(t)|^2=|\langle \downarrow^-|\langle -\alpha_{\rm sp}(g_1)| e^{-it H(g_2=0)} |\alpha_{\rm sp}(g_1)\rangle |\downarrow^+\rangle|^2+|\langle \downarrow^+|\langle \alpha_{\rm sp}(g_1)| e^{-it H(g_2=0)} |\alpha_{\rm sp}(g_1)\rangle |\downarrow^+\rangle|^2.
  \end{align}
Since $H(g_2)=\eta\omega\sigma_z/2+\omega_0\adaga$, the spin degree of freedom averages out in the $\eta\rightarrow \infty$ limit, while the coherent state simply acquires a time-dependent amplitude $\alpha_{\rm sp}(g_1,t)=\alpha_{\rm sp}(g_1)e^{-i\omega_0 t}$. Thus, Eq.~\eqref{eq:Lta} further simplifies to
\begin{align}\label{eq:L2}
|L(t)|^2=e^{-2|\alpha_{\rm sp}(g_1)|^2(1+\cos\omega_0t)}+e^{-2|\alpha_{\rm sp}(g_1)|^2(1-\cos\omega_0t)}.
  \end{align}
As $\alpha_{\rm sp}(g_1>1)\propto \sqrt{\eta}$, we can obtain the rate function in the $\eta\rightarrow\infty$ limit,
\begin{align}
r^{\infty}(t)\equiv \lim_{\eta\rightarrow\infty}r(t)=-\lim_{\eta\rightarrow\infty}\frac{1}{\eta}\log|L(t)|^2=\min_{\rm q=\pm} f_{q}(t)
  \end{align}
where $f_{\pm}(t)=2\alpha^2(g_1)(1\mp \cos(\omega_0t))$ with $\alpha(g_1)=\alpha_{\rm sp}(g_1)/\sqrt{\eta}=\sqrt{(g_1^2-g_1^{-2})/4}$ independent of $\eta$, and thus $f_{\pm}(t)$ a $\eta$-intensive function. At the times for which $f_{+}(t_c)=f_{-}(t_c)$ the rate function will display a kink, i.e., a non-analytical point, which here corresponds to $\omega_0t_c=\pi/2+n\pi$ with $n=0,1,\ldots$, and the rate function at those values amounts to $r^\infty(t_c)=2\alpha^2(g_1)=(g_1^2-g_1^{-2})/2$. This is shown in Fig. 3(c) of the main text.

A simple calculation allows us to obtain the critical exponent $\beta$ of the DPT-II, defined as $r^\infty(t_c)-r^\infty(t)\sim |t-t_c|^\beta$ with $|t-t_c|\ll 1$,  which follows from a Taylor expansion around $t_c$. This trivially leads to
\begin{align}
r^\infty(t_c)-r^\infty(t)=2\alpha^2(g_1)|t-t_c|
  \end{align}
so that the critical exponent is $\beta=1$, as given in the main text, and shown in Fig. 3(d). 

In addition, a further inspection of Eq.~\eqref{eq:L2} allows us to the finite-$\eta$ correction to $r^\infty(t_c)$. In particular, upon a substituting $\omega_0t_c=\pi/2+n\pi$ into $r(t)$ for a $\eta$ value, we obtain $r(t_c)=2\alpha^2(g_1)-\frac{\log 2}{\eta}$, as commented in the main text.





\end{widetext}


\end{document}